\documentclass[]{aiaa-tc}%  Info: http://www.ctan.org/tex-archive/macros/latex/contrib/aiaa/

\usepackage{lettrine}%  Dropped capital letter at beginning of paragraph
\usepackage{iepc2024}%  IEPC style file, adds formatting between the title and the authors, and in the footnote.  Requires the input \IEPCsubmissionnumber{xxx}.
\usepackage{amsmath}
\usepackage{physics}
\usepackage{fancyref}
\usepackage{caption}
\usepackage{subcaption}
\usepackage{url}

\DeclareMathOperator*{\argmin}{argmin}

\IEPCsubmissionnumber{138}%

\title{Time-Resolved Data-Driven Surrogates of Hall-effect Thrusters}%\thanks{Distribution Statement A: Approved for Public Release; Distribution is Unlimited. PA\# AFRL-2024-2990}
\author{
Adrian S. Wong\thanks{Computational Scientist, adrian.wong.ctr@us.af.mil\\Distribution Statement A: Approved for Public Release; Distribution is Unlimited. PA\# AFRL-2024-2990}\\%
{\normalsize\it{Jacobs Technology Group, Air Force Research Laboratory, Edwards AFB, California, 93524, USA}}\\%, Tullahoma, Tennessee, 37388, USA}}\\%
\\%
Christine M. Greve and Daniel Q. Eckhardt\\%
{\normalsize\it{In-space Propulsion Branch, Air Force Research Laboratory, Edwards AFB, California, 93524, USA}}\\%
\\%
}
\begin{document}
\maketitle

\begin{abstract}

Abstract: The treatment of Hall-effect thrusters as nonlinear, dynamical systems has emerged as a new perspective to understand and analyze data acquired from the thrusters. The acquisition of high-speed data that can resolve the characteristic high-frequency oscillations of these thruster enables additional levels of classification in these thrusters. Notably, these signals may serve as unique indicators for the full state of the system that can aid digital representations of thrusters and predictions of thruster dynamics. In this work, a Reservoir Computing framework is explored to build surrogate models from experimental time-series measurements of a Hall-effect thruster. Such a framework has shown immense promise for predicting the behavior of low-dimensional yet chaotic dynamical systems. In particular, the surrogates created by the Reservoir Computing framework are capable of both predicting the observed behavior of the thruster and estimating the values of certain measurements from others, known as inference.
\end{abstract}

\section*{Nomenclature}

\begin{center}
\begin{tabular}{ l l l l l}
$N$         & = number of nodes/neurons 		&	& $\hat g$	& = autonomous reservoir dynamics \\
$M$         & = number of time slices 			&	& $\eta$	& = Gaussian noise \\
$D$         & = number of inputs 				&	& $t$       & = time \\
$A$         & = adjacency matrix  				&	& $\Delta t$& = integrator time step \\
$B$         & = coupling matrix 	 			&	& $x$       & = state variables \\
$W$         & = weight matrix 	 				&	& $y$       & = measurement variables \\
$\lambda$	& = regularization parameter		&	& $\hat y$  & = estimated measurements \\
$\rho$      & = spectral radius 	 			&	& $r$       & = reservoir variables \\
$f$ 		& = unknown dynamical system 	 	&	& $z$    	& = inferred measurements \\
$h$ 		& = measurement function 	 		&	& $\tau$	& = time-delay\\
$g$			& = driven reservoir dynamics  		&	& &
\end{tabular}
\end{center}

\newpage

\section*{Declarations}

\subsection*{Funding}
This research was supported by the Air Force Office of Scientific Research under FA9550-23RQCOR001.

\subsection*{Conflicts of Interest}
The authors have no conflicts of interests to declare that are relevant to the content of this article.

\subsection*{Data Availability}
Data used in this work are available from the corresponding author on reasonable request, evaluated on a case-by-case basis.

\subsection*{Code Availability}
A general version of the Reservoir Computing code is available in the following repository: \url{https://github.com/adrianskw/libRC}.

\subsection*{Authors' Contributions}
Following the CRediT taxonomy, writing of the original draft, including conceptualization, methodology, investigation, and formal analysis was performed by Adrian Wong. Review and editing was performed by Adrian Wong and Christine Greve. Data curation, funding acquisition, and supervision provided by Daniel Eckhardt.

\newpage

\section{Introduction}

Hall-effect Thrusters (HETs) continue to grow in on-orbit presence due to their efficient propellant usage for simple maneuvers like altitude adjustment for constellation flight (Starlink, Kuiper, etc). As such, the need to improve the understanding of the fundamental plasma characteristics that underlie thruster performance persists. Transient plasma dynamics have been studied extensively to understand HETs oscillations, particularly due to the availability of new time-resolved diagnostic techniques, with the goal of understanding details of electron turbulence\cite{Lafleur2016, Lafleur2016a, Hagelaar2003, Jorns2018,Faraji2023, Faraji2023a}. In parallel, there is a small but growing interest in the study of HET oscillations and the tight coupling between observable system oscillations and thruster performance\cite{Greve2019, Jardin2024, Hara2014, Eckhardt2019}.  

The growing acknowledgment in the HET community is that dynamics of the plasma discharge is centrally important to understand plasma characteristics and ultimately predicting thruster performance\cite{LuccaFabris2015, Durot2014, Lobbia2009, Roberts2022}. It is understood that plasma instabilities play a role in the underlying dynamical phenomena of these devices based on the existence of distinct operating regimes or modes of the thruster\cite{Boeuf1998}. The most notable of these oscillations is termed a `breathing mode' oscillation characterized by an amplitude on the order of its mean value and a frequency in the 10-40 kHz range \cite{Barral2009, Barral2006, Eckhardt2017, Hara2014}. These oscillations can affect the erosion rate of the thruster channel, the efficiency of the thruster, and even the thrust. Furthermore, recent work has determined that measurement signals from the thruster that capture these dynamics may be capable of providing a unique representation of the thruster operating mode\cite{Greve2024}.

To this end, the work in Lobbia (2009) to synchronize multiple sensor readings of such oscillations over a finite spatial domain has aided the confirmation that dynamics of discrete plasma ejection events are related to characteristic plasma circuit oscillations\cite{Lobbia2009}. Dale and Jorns (2021) demonstrated that time-resolved laser diagnostics could be coupled with modeling techniques to understand the evolution of plasma properties over a single breathing mode oscillation\cite{Dale2021}. Jorns (2018) also used symbolic regression techniques to create a steady-state data-driven model for anomalous electron transport. 

More recently, data-driven modeling techniques have sought to use the specific time-resolved dynamics of a HET to improve understanding of a particular phenomenon\cite{Araki2021, Eckhardt2019, Greve2019, Greve2021, Martin2019}. Eckhardt et. al. studied attractor reconstruction of HET dynamics to provide an accurate method for mapping between a single input and single output signal\cite{Eckhardt2019}. Greve et. al. (2019) furthered this technique by using HET dynamics to optimize analytical performance of a model\cite{Greve2019}. Following this, Greve and Marsh (2024) develop a heuristic for steady state thruster operation utilizing the uniqueness of the thruster discharge current\cite{Greve2024}.  Each application has increasingly demonstrated the significance of dynamics to the performance of the thrusters and opened the door for more interesting studies such as signal prediction based on a known input.

In this work, we explore a data-driven technique to predict future time-resolved dynamics of a HET from measurement signals alone, specifically the Reservoir Computer (RC) framework, which includes Echo State Networks and Liquid State Machines\cite{Jaeger2001, Natschlaeger2002, Maass2002}. These systems are characterized by their random initiation as recurrent neural network, before `converting' to a single hidden-layer feedforward neural network. RCs also generally use a linear readout layer since it is the fastest and most efficient way of training the network. Standard usage of the RC framework can be conceptually divided into 3 phases, the listening, training, and predicting phase. 
%We give a brief outline here, deferring the detailed explanation for the later sections. The RC running as a recurrent neural network, driven by some input data, constitutes the listening phase. This is immediately followed by the training phase to acquire the readout weights of the network. An autonomous system can then be defined using the acquired weights, and the evolution of this autonomous system constitutes the predicting phase. 

The standout strength of a RC is its ability to quickly create a surrogate model that can predict future behavior of complex dynamical system, even in the presence of chaotic behavior\cite{Pathak2018, Platt2021}. It does so without any knowledge of the underlying system generating the data, with only mild assumptions on continuity and boundedness of the system\cite{Grigoryeva2021}. For low-dimensional nonlinear systems, the prediction performance of RCs are unmatched within the purely-data driven category compared to other neural network approaches\cite{Chattopadhyay2020, Shahi2022}. In this work, we treat the measured HET system as a low-dimensional nonlinear systems, where the measurement signal are tightly coupled via the plasma. We show that it is possible to predict future time-resolved dynamics of the system from historic measurements alone.

The fundamentals of RC and its behavior can be defined in a rigorous manner and is actively under study \cite{Grigoryeva2018, Grigoryeva2021, Hart2020, Hart2023, Hart2021, Yildiz2012, Zhang2011}. RCs have also seen many applications since its inception, such as reproducing of Lyapunov exponents, reconstructing attractors, determining causality, and being a \textit{state observer} \cite{Pathak2017, Grigoryeva2023, Lu2017, Lu2018, Weng2019, Huang2020}. The ability to \textit{observe} states means that the RC can estimate the state of the measured system from measurements. We will use the term \textit{infer} instead of \textit{observe} for the process of estimating across variables, rather than estimating the underlying state. This is specific to our use case since only measurements of the underlying system are available. We apply this inference capability to estimate all other measurements from the anode discharge current, with promising results. Our efforts demonstrate that working with time-resolved dynamics is plausible alternative, particularly when aided by a data-driven framework.

%The listening phase has been linked to the synchronization of chaos, which was very active just before the turn of the millennium \cite{Pecora1997, Pecora1990, Rulkov1995, Abarbanel1996, Kocarev1996}. The synchronization of the listening phase has been studied in some depth, but the synchronization in the inference phase has not been studied in great depth\cite{Hart2023, Grigoryeva2021, Hart2020, Lu2020}. 

\section{Data Preparation}

The experimental time-series data is of a HET placed in a large metallic cage, itself in a larger vacuum chamber \cite{MacDonald2016}. The HET is in a breathing-mode oscillation and the data set consists of roughly 125,000 sequential measurements, corresponding to roughly 80 breathing modes in total. The measurements comes from 9 sensors and were subject to scaling so that the data of each sensor lie within the interval $[-1,1]$. This is an important step to ensure that the behaviors of each variables is equally considered.

%Two of these sensors are the cathode and anode current. The remaining Beam Dump, Backwall, and 5 rings.

The data is first down-sampled such that every 5th data point is used, i.e. a 1:5 ratio. Various sampling ratios of 1:1, 1:2, and 1:10 were briefly tested but all produced similar results. The ratio of 1:5, corresponding to approximately 300 samples per breathing mode, was ultimately chosen to reduce computational burden. 

After down-sampling, the data was then divided into two partitions. The first partition consists of the first $50\%$ of the time-series and was used for the training the surrogate model. This is called the \textit{measurement window} and is described as $t_1 \le t \le t_M$. The remaining and latter $50\%$ of the time-series is used to verifying or test the trained surrogate model. This is called the \textit{prediction window} and is described as $t > t_M$ without specifying the terminal time. Various partitions were chosen between $50/50$ to $80/20$, all with similar results. The $50/50$ ratio was ultimately chosen so that the partitions were of equal size.

\section{Problem Statement and Proposed Approach}

Consider the situation where a vector time-series $y(t)=[y_1(t),y_2(t), \cdots, y_D(t)]^\top$ is given at discrete uniform intervals $t=\{ t_1, t_2, \cdots, t_M\}$ in some time-window. There are $D$ variables being measured at every instance of time, with $M$ time slices. Future predictions of these $D$ measurements are desired for times $t>t_M$, but no further information regarding the system is given.

The assumption is that $y$ are measurements of a deterministic dynamical system. Let this system have state variables $x$ and evolve according to some autonomous velocity field $\dot x=f(x)$. The measurements $y = h(x)+\eta$ occur at uniform intervals and are given by the measurement function $h$ acting on the state $x$. Additive noise $\eta$ (independent identically distributed) is also present. The system dynamics and its measurements can be formulated as follows.
\begin{equation}
\begin{split}
	\dot x&=f(x)\\
	y&=h(x)+\eta
\end{split}\label{eq:dyn}
\end{equation}

Neither the state variables $x$, its dynamics $\dot x=f(x)$, nor the measurement function $h$ are known. Predicting future measurements can be achieved by generating estimates $\hat y(t)$ for $t>t_M$ beyond the time-window of available measurements. Due to the availability of data but the opaqueness of the underlying system in \eqref{eq:dyn}, a purely data-driven surrogate model is necessary. 

To arrive at meaningful and verifiable results, a surrogate model must operate in the space of measurements $y$, not the state space $x$. This model must, at a minimum, recreate the measurements in the measurement window, which is verified using the first partition of data. A well-performing surrogate model should also generate predictions $\hat y$ that closely match the real values of $y$ in the short-term horizon. The prediction performance can be quantified and verified with the second partition of data.

In this work, we propose and consider the Reservoir Computing framework for training a shallow feedforward neural network as a surrogate model\cite{Jaeger2001,Maass2002,Pathak2018}. The trained network has one hidden layer and one readout layer, where the training involves only a linear least squares fit of the network trajectory onto the measurement data\cite{Schrauwen2007}. 

The shallow network architecture is much smaller than the deep neural network counterparts, and the requisite training only involves a non-iterative linear solve, which is incredibly fast\cite{Weng2019,Schrauwen2007}. The RC framework yields networks that are small in size (both width and depth) with very efficient learning, in terms of the number of training samples and the time to train. These factors combined make the RC framework incredibly appealing for developing surrogate models. 

\section{Reservoir Computing}

In this section, we will outline the structure of the RC framework, conceptually divided into four phases: listening, training, predicting, and inferring. The setup of the RC framework, such as the necessary preparation and specific constants for implementation, will be discussed in the next section. After the data $y$ has been prepared, it is used as an input signal to drive a non-autonomous dynamical system $r$, which has $N$ nodes, defined by the ordinary differential equation (ODE) below. 
\begin{equation}
\begin{split}
	\dot r & = -r+\tanh(Ar+By)
\end{split}\label{eq:rc1}
\end{equation}

This system $r$ can also be viewed as a recurrent neural network but, in the context of the RC framework, is called a \textit{reservoir}. The adjacency matrix $A$ describes how the nodes of the reservoir interact with one another, and the coupling matrix $B$ describes how the inputs $y$ are coupled to each node. Both $A$ and $B$ are randomly generated.

The system is numerically integrated forward in time with initial condition $r(0)$ allowed to be arbitrary, as the state of the reservoir converges exponentially to a unique state and maintains itself on that unique trajectory. This has been mathematically proven in many previous works. The generation of $r(t)$ using $y(t)$ is commonly called the \textit{listening phase}.
\begin{figure}
\centering
\begin{minipage}{.32\textwidth}
  \centering
  \includegraphics[width=\linewidth]{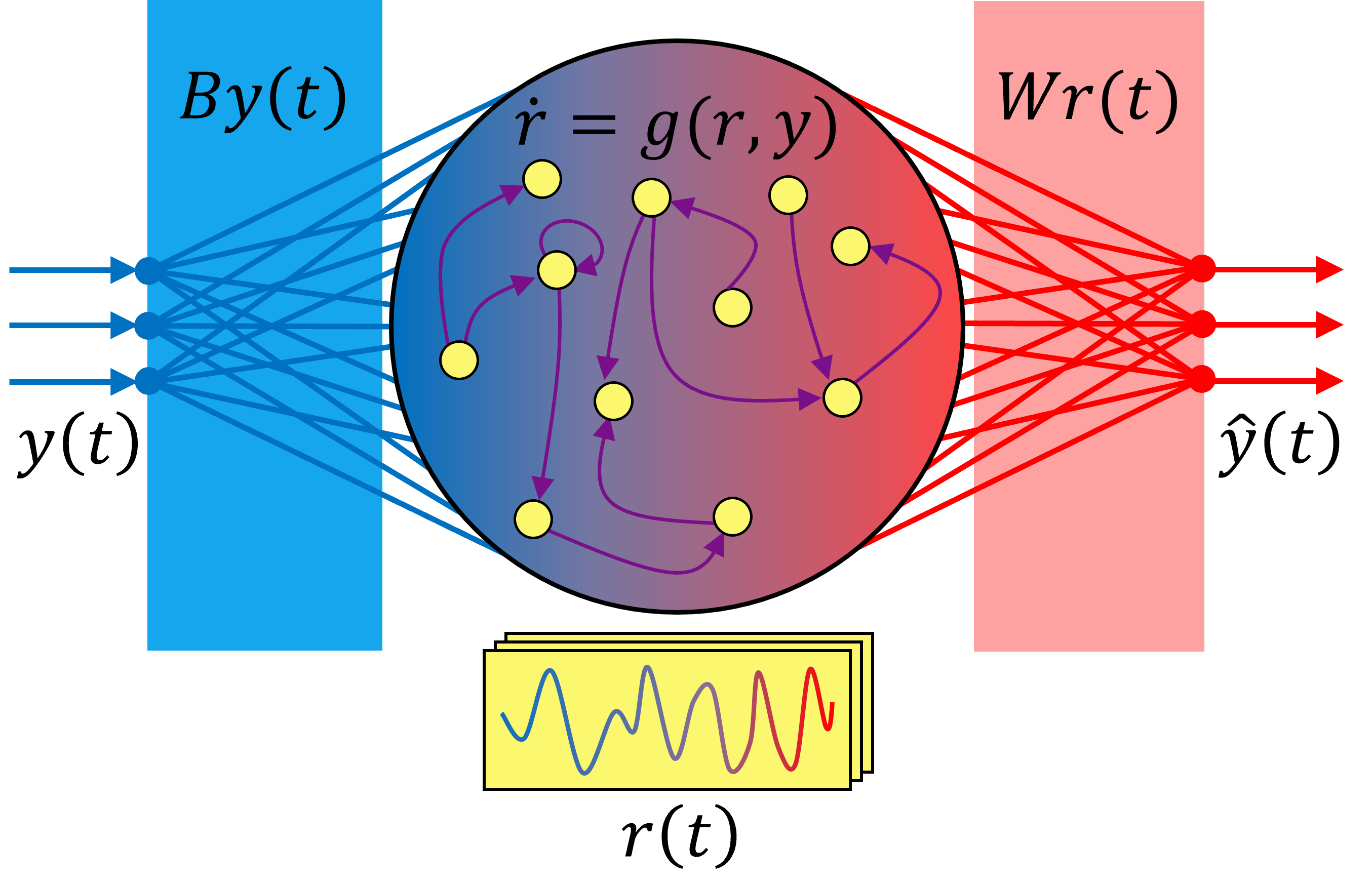}
  \captionof{figure}{Reservoir in the listening and training phase. The system $g(r,y) = -r+\tanh(Ar+By)$ is driven by the inputs $y$.}
  \label{fig:listening}
\end{minipage}%
\hfill
\begin{minipage}{.32\textwidth}
  \centering
  \includegraphics[width=\linewidth]{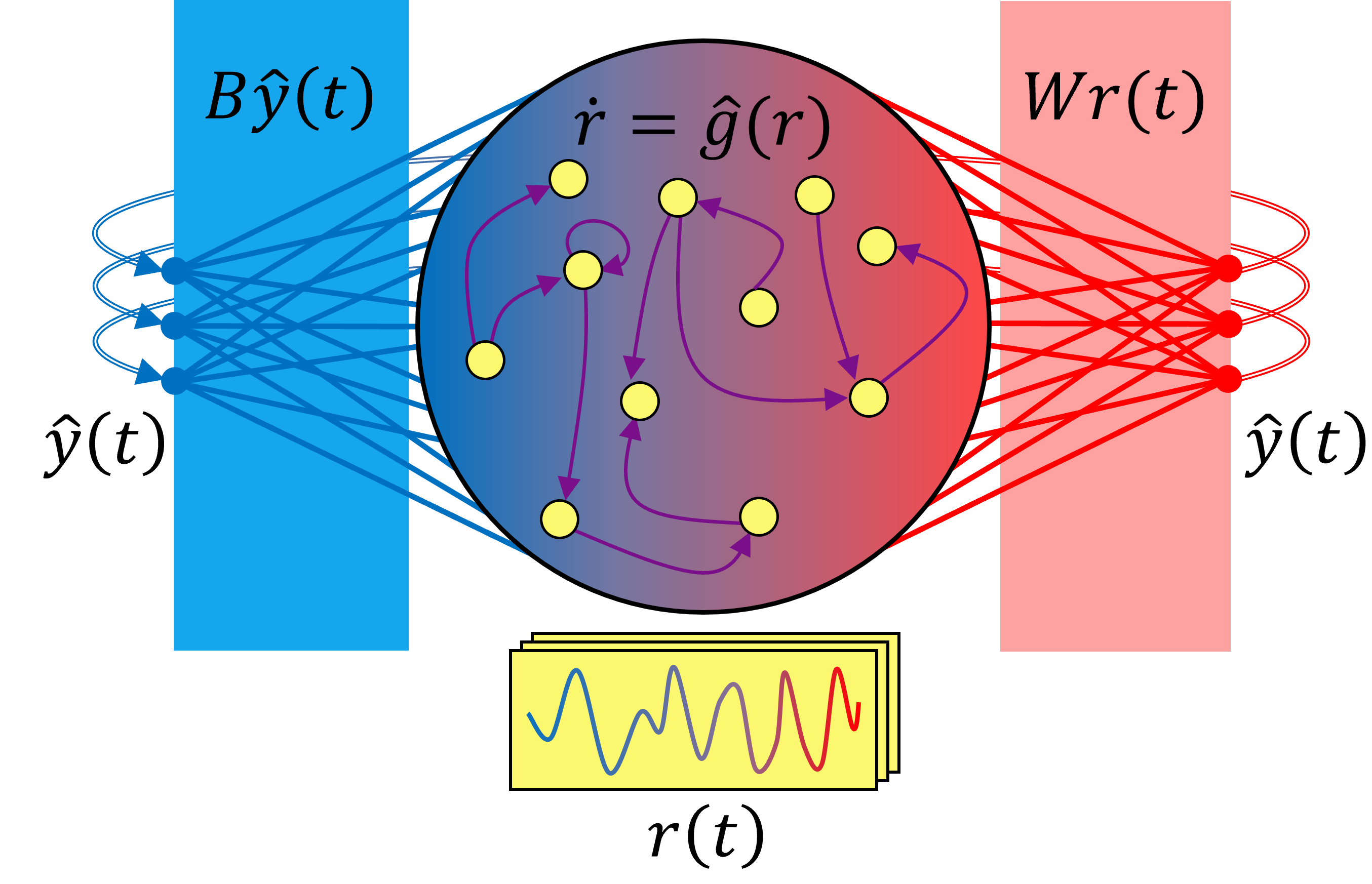}
  \captionof{figure}{Reservoir in the predicting phase. The system $\hat g(r) = -r+\tanh([A+BW]r)$ is autonomous since $\hat y(t) \equiv Wr(t)$.}
  \label{fig:predicting}
\end{minipage}
\hfill
\begin{minipage}{.32\textwidth}
  \centering
  \includegraphics[width=\linewidth]{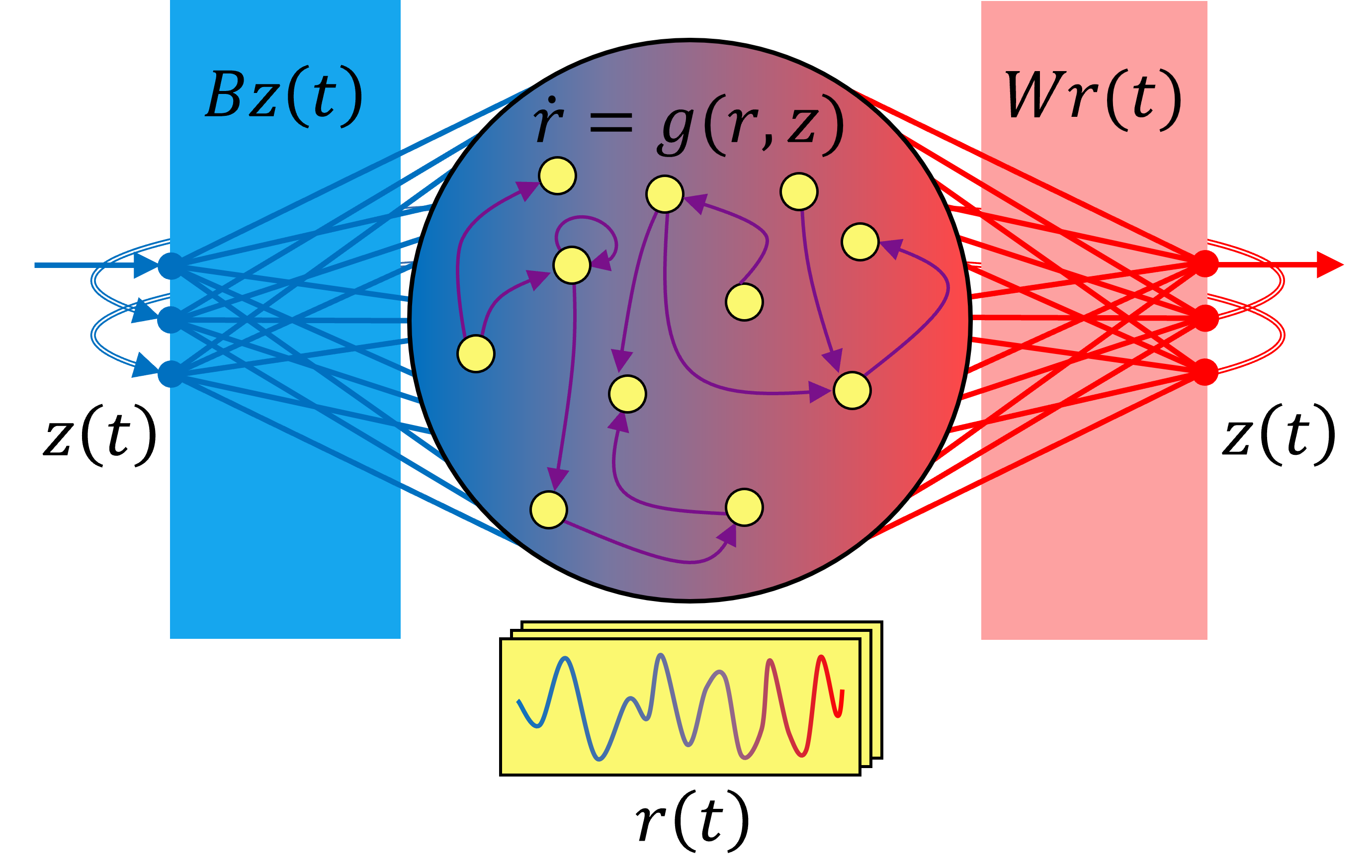}
  \captionof{figure}{Reservoir in the inferring phase. The system $g(r) = -r+\tanh(Ar+Bz)$ is driven by $z(t)\equiv[y_1(t),\hat y_2(t),\hat y_3(t)]^\top$.}
  \label{fig:inferring}
\end{minipage}
\end{figure}

There are a total of two trajectories to consider now -- the data $y(t)$ and the reservoir $r(t)$. The next step is to project or fit the reservoir trajectory $r(t)$ down onto the data $y(t)$ such that an error is minimized. Though more complicated types of fit can be used, this work makes exclusive use of the linear least-squares fit due to its speed and simplicity\cite{Schrauwen2007,Hart2021}. The cost function associated with this is the time-averaged Euclidean error. The minimization of this cost function is called the \textit{learning phase}.
\begin{equation}
\begin{split}
	y \simeq \hat y = Wr \Longleftrightarrow \argmin_{W} \sum_{m=1}^M \norm{y(t_m)-Wr(t_m)}
\end{split}\label{eq:fit}
\end{equation}

Due to the fact that the least squares fit will always contain a small amount of error, it becomes necessary to distinguish between the original data $y(t)$ and the attempted reconstruction. The fitted trajectory will be called $\hat y(t) \equiv Wr(t)$ and we can expect the individual states of the original and reconstruction to be numerically close in value, i.e. $ y \simeq \hat y$, and can be verified to be the case within the measurement window. This justifies the substitution $y \leftarrow \hat y$ in the prediction window, applied to \eqref{eq:rc1}, effectively using the outputs of the reservoir as its inputs. The result is the following autonomous reservoir and readout.
\begin{equation}
\begin{split}
	\dot r & = -r+\tanh([A+BW]r)\\
	\hat y & = Wr
\end{split}\label{eq:rc2}
\end{equation}

Integrating the autonomous reservoir above in the prediction window $t>t_M$, then retrieving the corresponding $\hat y(t)$ is called the \textit{predicting phase}. The initial condition in the prediction window is the terminal state $r(t_M)$ of the measurement window. We are now able to generate $\hat y$ for $t>t_M$, which are predictions of the measurement data beyond anything that the reservoir has been presented before. These predictions can then be compared to the data $y(t)$ in the prediction window. 

After an autonomous reservoir is obtained, define $z$ as a combination of measured and reconstructed variables. Consider the illustrative example corresponding to Figures \ref{fig:listening}, \ref{fig:predicting}, and \ref{fig:inferring}. There are 3 variables -- $y_1$ is the measured variable; $\hat y_2$ and $\hat y_3$ are the reconstructed variables. This specific structure is just an example -- there is is no restriction that data be input through the first variable, nor does it restrict the number of variables being used for inference\cite{Lu2017}. These dynamics are describes by the following equations, and is a form of Generalized Synchronization\cite{Pecora1997,Pecora1990}.
\begin{equation}
\begin{split}
		 z &= [y_1,\hat y_2,\hat y_3]^\top \\
	\dot r & = -r+\tanh(Ar+Bz)\\
		 z & = Wr
\end{split}\label{eq:rc3}
\end{equation}

We will describe the reservoir in this manner as the \textit{inferring phase}. It is very closely related to the predicting phase of \eqref{eq:rc2}, but with $y_1$ being presented to the trained reservoir. In comparison, the predicting phase does not have any data or measurements being presented to it at all. This can be interpreted as $\hat y_2$ and $\hat y_3$ being inferred from $y_1$, hence the name. The inferred trajectory $z(t)$ can be compared with the data $y(t)$ for $t>t_M$ in the same way as the predictions.

\section{Implementation Details}

The implementation of the RC framework involve many parameters, which leaves a lot of freedom for its users. We will specify our implementation in detail and, for context, the standard best-practices for implementing RC. Equation \eqref{eq:rc1} is a common variety of a Hopfield network, widely studied in the context or both biological and artificial neural networks. The $N\times N$ matrix $A$ is usually randomly generated and sparse. We use a total of $N=400$ nodes in our implementation. 

The sparsity of $A$ is the fraction of its elements being non-zero values. Our implementation has $2\%$ of the elements of $A$ randomly populated, with values picked randomly from a uniform distribution in the range $[-1,1]$. The remaining $98\%$ of entries are left with a value of zero. $2\%$ is a very common choice for the sparsity, and greater or lesser values tend toward bad predictions. Different types of distributions for assigning the non-zero values of $A$ do not result in significant behaviors. 

The spectral radius $\rho(A)$ (largest magnitude eigenvalue) of $A$ is usually prescribed such that $\rho(A)<1$. This is necessary for important stability criteria, allowing indifference to the initial condition $r(0)$ \cite{Yildiz2012,Zhang2011,Buehner2006}. This condition is enforced by randomly generating $A$ first, then scaling the entire matrix to achieve the desired $\rho(A)$. Values of $\rho(A)$ in the neighborhood of $1$ have demonstrated better predictions; we use a value of $\rho(A)=0.95$.

Also specific to the RC framework, the $N\times D$ matrix $B$ is usually randomly generated. Since we have data from 9 sensors, we have that $D=9$. This work uses a relatively sparse $B$ such that each $N$ reservoir nodes is driven by exactly one of the $D$ inputs, with a coupling strength also randomly generated from the uniform distribution with range $[-1,1]$. The outcome is that among the $ND$ entries of $B$, $N$ of them are populated.

We used the 2nd-order Runge-Kutta (RK2) numerical scheme to integrate the ODE \eqref{eq:rc1} and generate the reservoir trajectory $r(t)$ with time step $\Delta t=0.22$. For reasonable time steps, a higher or lowest integration schemes has minimal to no effect on the generated trajectory. However, increases and decreases in $\Delta t$ result in predictions with higher and lower frequencies respectively. For breathing mode oscillations, $\Delta t$ is straightforward to adjust to achieve desired behavior.

The training phase is accomplished by minimizing the cost function \eqref{eq:fit}. As a practical matter, this is most efficiently accomplished by matrix inversion. Define a $D\times M$ matrix $Y$ as a representation of the vector time-series $y(t)$ where each column of $Y$ is the measured variables and consecutive columns being consecutive instances in time. Also define a $N\times M$ matrix $R$ as a representation of the vector time-series $r(t)$ in the same manner. Notice that $W$ should ideally satisfy $Y = WR$. Solving for the non-square matrix $W$ can be achieved via matrix multiplication and computing an inverse.
\begin{equation}
\begin{split}
	Y &= WR \\
	YR^\top &= WRR^\top\\
	YR^\top (RR^\top)^{-1} &= W
\end{split}
\end{equation}

However, the condition number of the $(RR^\top)$ matrix is usually large, so solving for the inverse is numerically ill-posed. \textit{Tikhonov regularization} is an attempt to alleviate this issue by modifying the problem slightly such that the inverse of the matrix $(RR^\top+\lambda I)$ is evaluated instead, with $\lambda$ being a small regularization constant and $I$ being the appropriately sized identity matrix. Intuitively, this $\lambda$ is the degree to which the measurement noise is ignored, effectively acting as the attenuation strength of a low-pass filter. We use a value of $\lambda=0.01$. The $D\times N$ matrix $W$ is evaluated as follows.
\begin{equation}
\begin{split}
	W = YR^\top (RR^\top+\lambda I)^{-1}
\end{split}\label{eq:matrix}
\end{equation}

Lastly, for the inference phase, current measurements of the cathode (\textit{Cathode} in Figure \ref{fig:all-in-one}) was used as the input for the inference process, with all other variables being inferred. This corresponds to \textit{Cathode} being used as $y_1$ in the example of \eqref{eq:rc3}. 

\section{Results and Discussion}

\begin{figure}
\centering
\includegraphics[width=\linewidth]{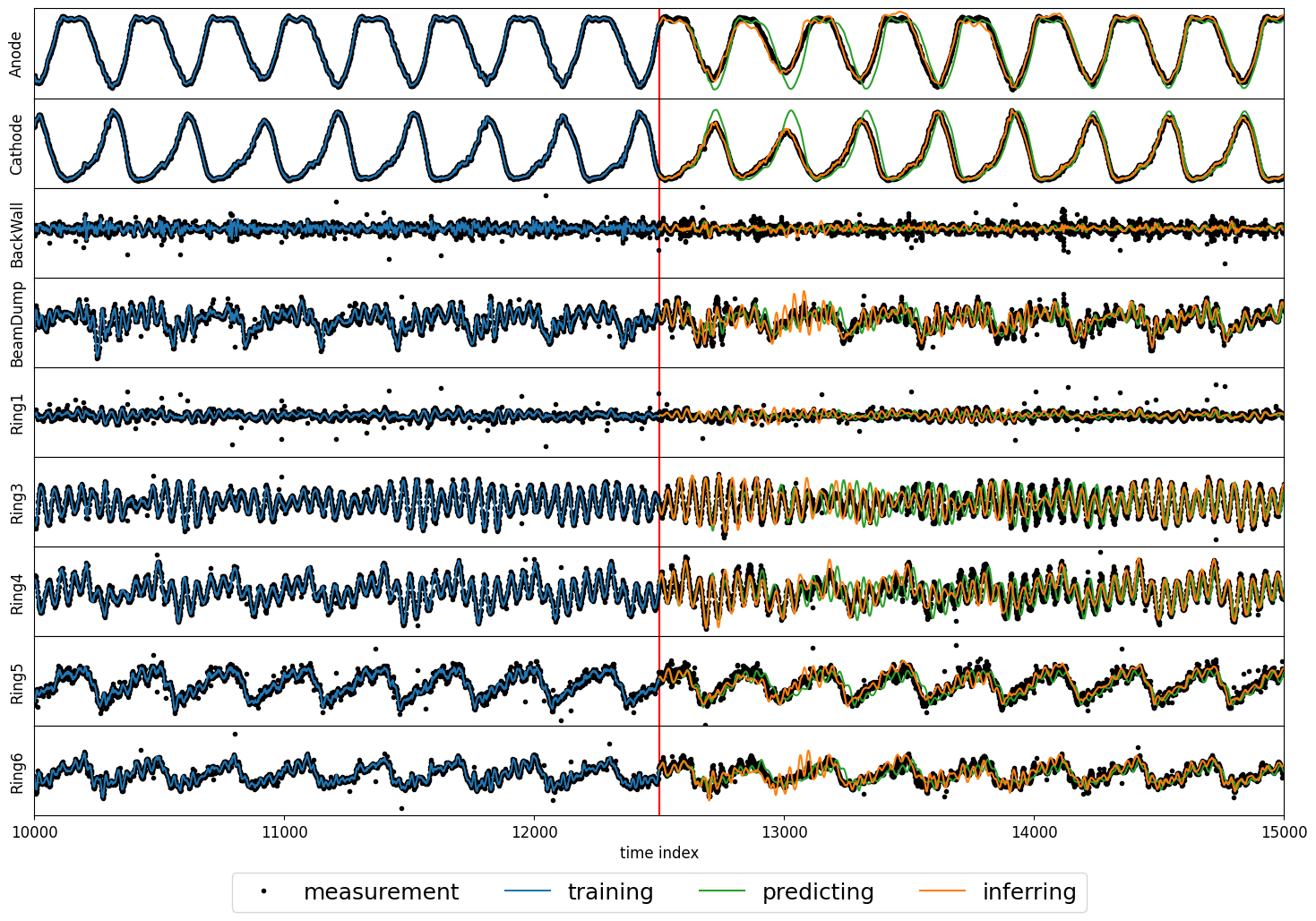}
\caption{The various outputs of the surrogate model are compared with the measurement data in black, with a red line separating the measurement and prediction windows. The left half shows the output of the training phase \eqref{eq:fit} in blue. The right half shows the prediction \eqref{eq:rc2} in green and the inference \eqref{eq:rc3} in orange. All plots have ranges roughly between $[-1,1]$ due to the initial re-scaling of the data.}\label{fig:all-in-one}
\end{figure}

The output of the training, predicting, and inferring phases are compared to the measurement data in Figure \ref{fig:all-in-one}. Only a fraction of the time series is plotted to avoid clutter. The red line lies on the boundary between the measurement window $0<t<t_M$ and prediction window $t>t_M$. 

The measurement data is the ground truth upon which all the other signals are compared with. As shown in the left half of Figure \ref{fig:all-in-one}, the time-series output of the training phase closely approximates the target signal except for the sharpest peaks. This comes as no surprise because $r(t)$ is a much higher-dimensional signal than $y(t)$, allowing for more parameters of the linear fit $\hat y(t) = Wr(t)$, thus resulting in a good approximation. The sharp peaks are avoided due to the regularization constant $\lambda$ in \eqref{eq:matrix}.

In the left half of Figure \ref{fig:all-in-one}, both the predictions and the inference capture the behavior of the measurements well in the short term. As the systems ventures deeper into the prediction window, the predictions begin to deviate from the measurements slightly whereas the inference tracks the measurements comparatively better. The degree of deviation can be quantified using the RMSE defined in \eqref{eq:rmse} or the PC \eqref{eq:pc}. PC is defined in \eqref{eq:pc} succinctly using the covariance and standard deviation function.

PC is a slightly better comparison tool since it allows for comparisons across different signals, whereas the NMSE cannot due to the different scales of the signals. Other signals were tested in the inference process, and some cases where multiple signals were used as inputs. Usage of \textit{Cathode} alone provided the best results for inference in term of RMSE and PC. 

\begin{equation}
\begin{split}
	\text{RMSE}[y(t),\hat y(t)] = \sqrt{\sum_{d=1}^D \sum_{m=1}^M \abs{y_d(t_m)-\hat y_d(t_m)}^2}
\end{split}\label{eq:rmse}
\end{equation}

\begin{equation}
\begin{split}
	\text{PC}[y(t),\hat y(t)] =  \frac{\text{cov}[y(t),\hat y(t)]}{\text{stdev}\,y(t)\;\text{stdev}\,\hat y(t)}
\end{split}\label{eq:pc}
\end{equation}

\begin{figure}
\centering
\begin{minipage}{.48\textwidth}
  	\centering
	\includegraphics[width=\linewidth]{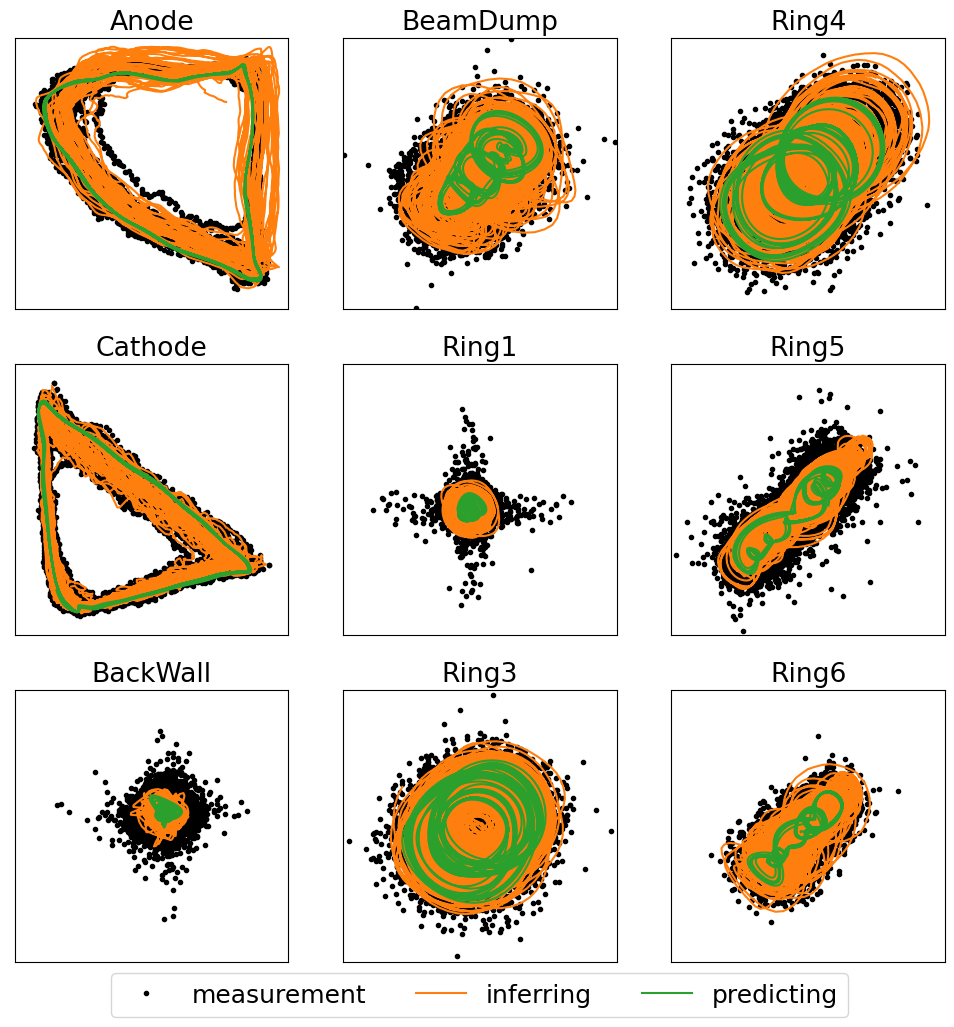}
	\caption{TLPPs of the variables of interest, with measurements in black, inference in orange, and prediction in green.}\label{fig:tlpp}
\end{minipage}%
\hfill
\begin{minipage}{.48\textwidth}
  	\centering
	\includegraphics[width=\linewidth]{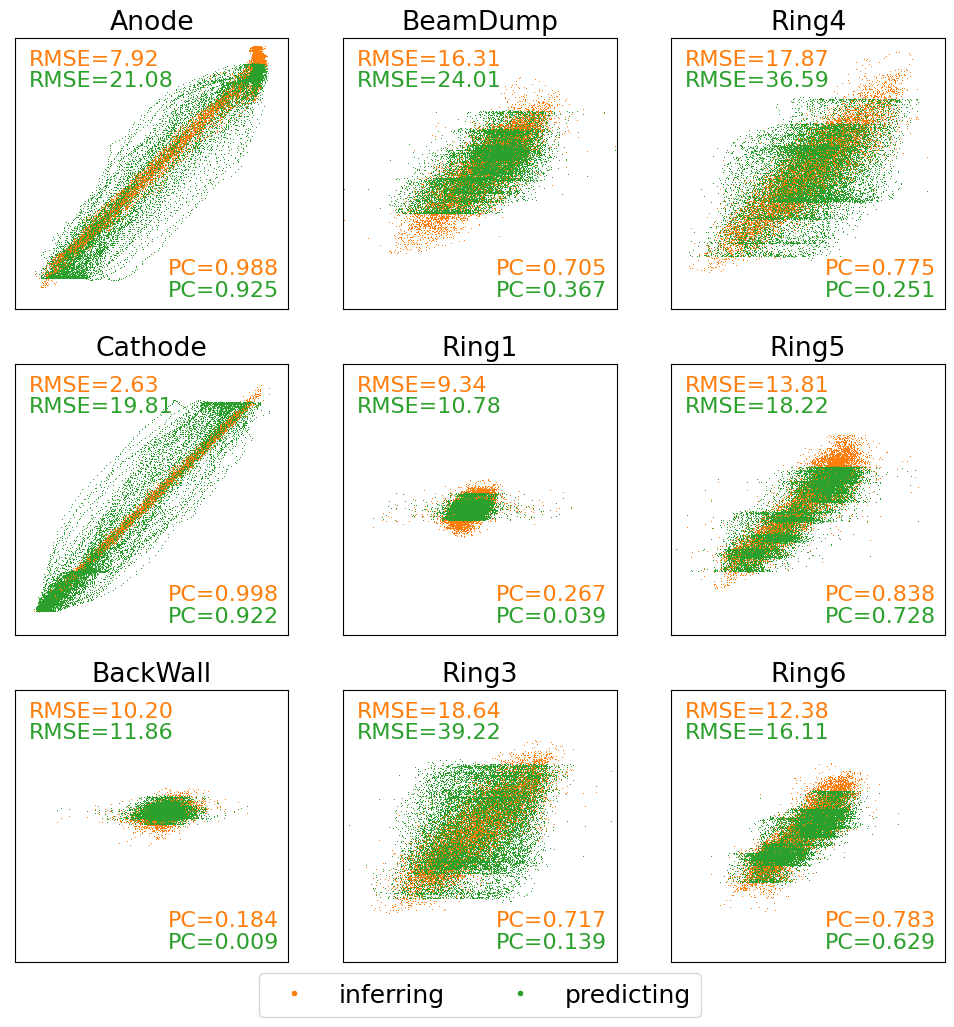}
	\caption{Synchronization plots of the variables of interest, along with the RMSE and PC for the predictions and inference respectively.}\label{fig:corr}
\end{minipage}
\end{figure}

Rather than simply comparing the time-series, there are other methods of visualization and comparison for these signals. The Time-Lag Phase Portrait (TLPP) of each phase, as shown in Figure \ref{fig:tlpp} for the prediction window, can also be compared to one another. The TLPP is a common technique in nonlienar time-series analysis, useful for comparing the long term behavior of the time series and producing informative yet condensed plots. In this figure, the TLPPs are time-series data plotted against their time-lagged value (i.e. $[y(t),y(t-\tau)]^\top$ with time-delay $\tau$), which effectively ``increases the dimension'' of the scalar signal. 

Another form of phase portraits commonly used in chaotic time-series analysis are shown in \ref{fig:corr}, where the variables are scalar signals of a reference and its comparison. In this work, we will call them \textit{synchronization plots}. The reference will be the measurements and the comparisons will be either the predictions or the inference , generating $[y(t),\hat y(t)]^\top$) and $[y(t),z(t)]^\top$) respectively. A comparison signal that reproduces the behavior of the reference signal will be tightly distributed along the diagonal, whereas signals that do not will be unstructured, generally resembling a ``ball of noise''.

The signals of \textit{Anode} and \textit{Cathode} are reconstructed well by both the predicting and inferring phase. The predicting phase tends to generate periodic signals as outputs, which is understandable given that the target signal is pseudo-periodic. The inferring phase, on the other hand, is able to capture the finer-scale behaviors of the systems, which explains both the higher PC and lower RMSE. It is also worth mentioning that the \textit{BackWall} and \textit{Ring1} signals have consistently low PC. Physical intuition suggest this is due to the place of the respective sensors that are up-stream from the thruster, resulting in signal that resemble noise. It is encouraging to see that neither the predicting phase nor inference phase attempts to reconstruct the noise in a detrimental manner. That is, feeding in `pure noise' has not resulted in bad reconstructions of the other signals.

There is a patch in the prediction window, not long after the measurement window, where the predicting phase consistently deviates from the measurements of the system. The inferring phase, most likely due to the presence of data, is able to overcome this deviation. The reconstruction also consistently ignores the noisiest parts of the signals, but it is unclear exactly how the denoising ability of the RC works. A noteworthy result is that the reconstruction of the inferring phase is consistently better than the prediction phase, as measured by higher PC and lower RMSE, for all measured signals. This is encouraging as it suggests that input to the trained surrogate can be used to improve and maintain estimates of the system, which suggests a possible mechanism for developing Digital Twins. This indicated a substantial advantage of using data to maintain the synchronization between the reservoir and the thruster, as compared to not using data at all. 

\begin{figure}
\centering
\end{figure}

\begin{table}[]
\centering
\begin{tabular}{||c|c|c|c|c||} 
 \hline
 Variable & Prediction PC & Inference PC & Prediction RMSE & Inference RMSE\\ [0.5ex] 
 \hline\hline
 Anode    	& 0.925 & 0.988 & 21.08	& 7.92 \\ 
 \hline
 Cathode  	& 0.922 & 0.998 & 19.81 & 2.63 \\
 \hline
 Backwall 	& 0.009 & 0.184 & 11.86	& 10.20 \\
 \hline
 BeamDump 	& 0.367 & 0.705 & 24.01 & 16.31 \\
 \hline
 Ring1 		& 0.039 & 0.267 & 10.78 & 9.34 \\ 
 \hline
 Ring3 		& 0.139 & 0.717 & 39.22 & 18.64 \\ 
 \hline
 Ring4 		& 0.251 & 0.775 & 36.59 & 17.87 \\ 
 \hline
 Ring5 		& 0.728 & 0.838 & 18.22 & 13.81 \\ 
 \hline
 Ring6 		& 0.629 & 0.783 & 16.11 & 12.38 \\ 
 \hline
\end{tabular}
\caption{Metrics of quality for all the measured variables considered.}
\end{table}

\section{Conclusion}

This work describes an RC framework for developing and utilizing a surrogate model for measurements of a HET system in a breathing mode, achieved in a purely data-driven manner. The development of the surrogate model is straightforward and simple, and the result is a dynamical forward model that is capable of producing predictions of measurements, surrogate to the underlying system. The quality of this surrogate model can be quantified by its ability to make predictions of future measurements, and it shows promising results matching the measurements in the predicting and inferring phases. 

The RC framework has a small memory footprint and performs well under linear readouts. Specifically, the small footprint refers to the relatively small number of weights required to describe the model, as compared to deep feedforward networks and other recurrent networks\cite{Chattopadhyay2020,Shahi2022}. As evidence, the surrogate used in this work has a total of 7209 floating-point weights, which is roughly the size of a 80-node fully-connected single-layer feedforward network. 

Unlike the RC surrogate models, other networks are generally not amenable with linear training either. Linear training is not iterative, involving only matrix multiplication and evaluation of a simple matrix inverse. Along with the small footprint, these two factors contribute to the impressive speed of the RC. The wall-time for the surrogate in the listening, predicting, and inferring phases are less than 2 seconds each, using Python on a standard desktop computer. The training phase alone took $0.16$ seconds. After including the listening phase, the entire training process is completed in less than 2 seconds.

There are two more attribute of this surrogate model that also stand out. The first of which is the relatively low number of samples that is required to train it, requiring only $10^4$ to $10^5$ input-output pairs, sometimes even much less, to produce good predictions and inference. The last outstanding property of these surrogate models, and perhaps the most interesting one, is the ability to continuously update the state of the model with incoming data. This was called the inferring phase, and it appears to be a plausible mechanism of developing a low-fidelity digital twins\cite{Kong2022, NationalAcademies2023}. 

There are, however, some caveats to the RC approach. It is fundamentally a phenomenological description of the observed data, without little to no knowledge of the relevant physics -- it is a low-fidelity model. Some attempts have been made to incorporate a detailed physics-based model to RC, similar to data assimilation techniques, but this first requires having some high-fidelity model\cite{Vlachas2020}.

Another caveat is that the roles of hyper-parameters are still not well understood. These are the regularization parameter $\lambda$, spectral radius $\rho(A)$, sparsity of $A$, and basically anything that informs the structure of $A$ and $B$. They need to be discovered or tuned manually on a case-by-case basis. Fortunately, the range of hyper-parameters that result in a well-performing RC is large, but the lack of rigorous understanding leaves the choice of hyper-parameters to seem arbitrary.

\section*{Acknowledgments}

The authors thank Robert Martin and Justin Koo for their guidance and insightful discussions during their time at the Air Force Research Laboratory.

\bibliographystyle{vancouver}
\bibliography{iepc2024}

%\section{Extra}
%
%
%Until recently, data-driven methods have been largely overlooked in favor of physically-derived methods. It was only until recently, following the last decade's explosion of machine learning research, that various data-driven methods matured, thus becoming a viable and attractive alternative to physically-derived methods. The RC framework is a data-driven method that has shown extraordinary promise in producing surrogate models of low-dimensional dynamical systems, even if the targeted system exhibits chaotic behavior. In this work, we apply the RC framework towards building surrogate models
%These data-driven methods are usually not derived from physical principles. Though there are ways of encoding physical constraints into these methods, they are generally viewed as low(er) fidelity methods because certain important physical properties of the target system are not guaranteed to hold true. Physically derived methods, on the other hand, have all the desired physics accounted for. To the extent that some physics is missing, these are usually intentionally omitted for the sake of tractability or simplification, making them high(er) fidelity methods. 
%All to all coupling. Work on a well defined attractor low dimensional
%Probe outage/signal dropout
%Diagnostics are complex and 
%LIF systems and Thompson scattering are hard to setup. Correlated readily available data to difficult to setup data.
%There are limitations of course. There's no guarantee 
%Does well with accuracy and training

\end{document}